\documentclass[12pt]{iopart}
\usepackage{times}
\usepackage{mathptmx}
\usepackage{graphicx}

\begin{document}

\title[Cyclotron Resonance in Two-Dimensional Electron \ldots]
{Cyclotron Resonance in Two-Dimensional Electron Single-Layers
and Double-Layers in Tilted Magnetic Field}
\author{N~Goncharuk, L~Smr\v{c}ka and J~Ku\v{c}era}
\address{Institute of Physics ASCR, Cukrovarnick\'a 10, 162 53 Praha
6, Czech Republic}

\begin{abstract}
The far-infrared absorption in two-dimensional electron layers subject 
to magnetic field of general orientation was studied theoretically.
The Kubo formula was employed to derive diagonal components of the
magneto-conductivity tensor of two-dimensional electron
single-layers and double-layers.  A parabolic quantum well was
used to model single-layer systems. Bilayer systems were
represented by a pair of tunnel-coupled, strictly two-dimensional
quantum wells. Obtained results were compared to experimental 
data. 
\end{abstract}

\section{Introduction}
\label{intro}

The absorption of far-infrared radiation in two-dimensional electron
layers subject to magnetic fields is used to determine the electron
cyclotron mass. In a {\em quasi-classical} picture of the cyclotron
resonance a magnetic field $B_{\perp}$, perpendicular to the plane of
the two-dimensional electron gas (2DEG), drives electrons with the cyclotron
frequency $\omega_c$ along trajectories similar to the shape of a
Fermi contour, but multiplied by $\hbar/(|e|B_{\perp})$, and rotated
by $\pi/2$.  The frequency $\omega$ of circularly polarized components
of the electromagnetic wave either adds to or subtracts from
$\omega_c$. The current $\vec{\mathcal J}(\omega)$ induced in the
layer by the electric field $\vec{\mathcal E}(\omega)$ dissipates, and
the corresponding power $P$ is expressed by
\begin{equation}
\label{power}
P= \frac{1}{T}\int_0^T\vec{\mathcal J}(\omega)\vec{\mathcal E}(\omega)
dt =\frac{1}{T} \sum_{\alpha
=x,y}\int_0^T\sigma_{\alpha\alpha}{\mathcal E}_{\alpha}^2(\omega) dt.
\end{equation}
In this expression $T=2\pi/\omega$ is a period of the cyclotron motion
and $\sigma_{\alpha\alpha}$ are the diagonal components of the
conductivity tensor. The absorption power $P$ reaches maximum 
at the {\em cyclotron resonance} frequency, i.e. when 
$\omega = \omega_c$.  The magnetic field at which the cyclotron 
resonance occurs, $B_{\perp c}$, defines the cyclotron effective 
mass by $m^*_c = |e|B_{\perp c}/\omega$. For circular Fermi contours 
the cyclotron effective mass and the effective mass are identical, 
$m^*_c = m^*$.  For the case of anisotropic effective mass 
$m^*_x \neq m^*_y$ (an elliptic Fermi line) $m^*_c = \sqrt{m^*_x m^*_y}$. 
In a {\em quantum-mechanical} picture, the electron states are quantized 
by the magnetic field $B_{\perp}$ into Landau levels and transitions 
between adjacent levels are measured.

Here we present the theoretical study of the cyclotron resonance in
symmetric 2D single-layers and double-layers subject to magnetic
fields of general orientation. In this case the in-plane cyclotron
mass becomes a function of both $B_{\|}$ and $B_{\perp}$.

A parabolic quantum well with a confining potential
$V_{conf}(z)=m^*\Omega^2z^2/2$ is used to model single-layer
systems. Bilayer systems are represented by a pair of tunnel-coupled,
strictly two-dimensional quantum wells, located at the distance
$d$. These simple models allow to carry out most calculations
analytically and are still able to capture the essence of physics of
single-layer and bilayer systems. Results obtained for a single well 
are compared to experimental data of Takaoka {\it et al}~\cite{takaoka}.

Assuming the Landau gauge of the vector potential,
$\vec{A}=(-B_{\perp}y +B_{\|}z, 0, 0)$, and a wave function in the
form $\psi \propto e^{ik_x x}\, \varphi(y,z)$, the Hamiltonian can
be written as
\begin{equation}
\label{hamilton}
H=\frac{1}{2m^*}(\hbar k_x +eB_{\perp}y-eB_{\|}z)^2+\frac{p_y^2}{2m^*}
+\frac{p_z^2}{2m^*}+V_{conf}(z).
\end{equation}
Its eigenenergies are Landau levels degenerated in $k_x$ with a factor
$|e|B_{\perp}/h$. For $B_{\|}=0$ the variables $y$ and $z$ in
the corresponding Schr\"odinger equation can be separated and for two
lowest electron subbands denoted by an index $j=0,1$ we get
two fans of Landau levels with the level index $i=0,1,2,\ldots$.

In tilted fields, the in-plane component $B_{\|}$ influences
strongly the orbital motion of electrons and modifies the structure of
Landau levels corresponding to $B_{\|}=0$. In our symmetric
systems a part of the Hamiltonian proportional to a product $yz$ is 
responsible for mixing of levels from different subbands when their 
level indeces $i$ differ by $\pm 1$.  We denote the new modified fans 
of Landau levels by a subband index $m=0,1$ and the new levels by an index
$n=0,1,2,\ldots$. Note that in our symmetric systems the
eigenfunctions $\varphi_{m,n}$ of the Hamiltonian are either even or odd:
\begin{equation}
\label{evenodd}
\varphi_{m,n}(y,z)=(-1)^{m+n} \varphi_{m,n}(-y,-z).
\end{equation}

\section{Magneto-conductivity tensor}
\label{conductivity}

According to (\ref{power}) only the ``in-phase'' parts of 
the diagonal components of the
magneto-conductivity tensor are  necessary to evaluate the dissipated
power.  We employed the Kubo formula to derive
$\sigma_{\alpha\alpha}$ of single-layers and double-layers subjected
to magnetic fields of general orientation. Since our  aim is to
study mainly the field-dependence of the energy levels
$E_{m,n}=E_{m,n}(B_{\|}, B_{\perp})$ and of the transitions
between them, we completely neglect the broadening of levels by
scattering of electrons. With this simplifying assumption we get, as a
response to the linearly polarized electromagnetic wave $\vec{\mathcal
E}(\vec{r},t) = 2\vec{\mathcal E}\cos(\vec{q}\vec{r}-\omega\, t)$ in the 
dipole approximation, the conductivity in the form
\begin{equation}
\label{kubo}
\sigma_{\alpha\alpha} = \frac{2\pi e^2}{\omega}\,\sum_{\mu,\nu} f(E_{\nu})
|\langle\mu|v_{\alpha}|\nu\rangle|^2
\left[\delta(E_{\nu}-E_{\mu}+\hbar \omega)-
\delta(E_{\nu}-E_{\mu}-\hbar \omega)\right].
\end{equation}
Here $f$ is the Fermi-Dirac distribution function, $\mu$ or $\nu$
stands for a Landau level double-index $m,n$ and $\alpha$ denotes $x$
and $y$. The in-plane components of the
conductivity $\sigma_{xx}$ and $\sigma_{yy}$ involve the matrix
elements of
\begin{equation}
\label{velocity}
 v_x  = -i\frac{\hbar}{m^*}\frac{\partial}{\partial x} -\omega_{\perp}y
+\omega_{\|}z, \;\;\;\;\;
 v_y   =  -i \frac{\hbar}{m^*}\frac{\partial}{\partial y}\;\;\; ,
\end{equation}
the in-plane components of the velocity operator.

The single-layer with a parabolic confining potential represents a
particularly simple example which can be solved analytically. In
tilted magnetic fields, the cyclotron frequency corresponding to the
perpendicular field configuration, $\omega_{\perp} =
|e|B_{\perp}/m^*$, and the frequency $\Omega$ defining the separation
of two subbands, are replaced by $\omega_1$ and $\omega_2$ given by
\begin{equation}
\omega_{1,2} =\frac{\sqrt{\omega_{\|}^2+(\omega_{\perp}+\Omega)^2}
\mp \sqrt{\omega_{\|}^2+(\omega_{\perp}-\Omega)^2}}{2} ,
\label{omega12}
\end{equation}
where $\omega_{\|}$ means $\omega_{\|} = |e|B_{\|}/m^*$.

The eigenenergies $E_{m,n}$ read
\begin{equation}
E_{m,n}=\hbar \omega_2(m+\frac{1}{2})+\hbar \omega_1(n+\frac{1}{2}).
\label{eigen}
\end{equation}

It follows from equations (\ref{kubo}) and (\ref{velocity}) that only
intrasubband transitions, $\delta m=0$, between the neighbouring
Landau levels, $\delta n=\pm 1$, or intersubband transitions, $\delta
m=\pm 1$, between the Landau levels with the same index $n$, $\delta
n=0$, are possible.  Two different values of the cyclotron effective
mass are connected with two processes, $m^*_{c,1} = |e|B_{\perp
c}/\omega_1$ and $m^*_{c,2} = |e|B_{\perp c}/\omega_2$. Moreover, the
conductivity can be  written as a linear combination of
\begin{eqnarray}
\label{parcond}
\sigma_1 & = & \frac{\pi e^2}{m^* \omega} \sum_{m,n} \hbar \omega_1(n+1)
[f(E_{m,n})-f(E_{m,n+1})][\delta(\hbar\omega_1-\hbar\omega)-
\delta(\hbar\omega_1+\hbar\omega)]  ,\\
\sigma_2 & = & \frac{\pi e^2}{m^* \omega} \sum_{m,n} \hbar \omega_2(m+1)
[f(E_{m,n})-f(E_{m+1,n})][\delta(\hbar\omega_2-\hbar\omega)-
\delta(\hbar\omega_2+\hbar\omega)]  . \nonumber
\end{eqnarray}
The components of the conductivity tensor read
\begin{equation}
\label{sigma}
\sigma_{\alpha\alpha} = W_{\alpha,1}\sigma_1 +  W_{\alpha,2}\sigma_2
\end{equation}
where the weight factors $W_{\alpha,1}$ and $W_{\alpha,2}$   
are different for $\sigma_{xx}$ and $\sigma_{yy}$
\begin{eqnarray}
\label{weights}
W_{x,1}= \frac{\omega_2^2-\omega_t^2}{\omega_2^2-\omega_1^2} , &\;\;\;\;  &
W_{x,2}= \frac{\omega_t^2-\omega_1^2}{\omega_2^2-\omega_1^2} ,\\
W_{y,1}=\frac{\omega_2^2}{\Omega^2}\, 
\frac{\Omega^2-\omega_1^2}{\omega_2^2-\omega_1^2} , &\;\;\;\;  &
W_{y,2}=\frac{\omega_1^2}{\Omega^2}\, 
\frac{\omega_2^2-\Omega^2}{\omega_2^2-\omega_1^2} ,\nonumber
\end{eqnarray}
and  $\omega_t$ is given by 
$\omega_t^2 = \omega_{\perp}^2 + \omega_{\|}^2$.

The detailed description of the model of the bilayer systems 
can be found e.g. in \cite{hu}. This model results in a pair 
of coupled differential equations for one-dimensional 
wave functions $\varphi_L(y)$ and 
$\varphi_R(y)$ localized in the left and right wells:
\begin{eqnarray}
\label{eqns}
-\frac{\hbar}{2m^*}\varphi_L^{''}(y) + 
\frac{m^* \omega_{\perp}^2}{2}\left(y+\frac{B_{\|}}
{B_{\perp}}\frac{d}{2}\right)^2 \varphi_L(y) 
+ t\; \varphi_R(y) & = & E \varphi_L(y) ,\\
-\frac{\hbar}{2m^*}\varphi_R^{''}(y) + 
\frac{m^* \omega_{\perp}^2}{2}\left(y-\frac{B_{\|}}
{B_{\perp}}\frac{d}{2}\right)^2 \varphi_R(y) + t\; \varphi_L(y)
& = & E \varphi_R(y).
\end{eqnarray}
Here, the orbit centres in individual wells are
shifted due to the in-plane component of the magnetic field and the
hopping integral $t$ describes the coupling of two two-dimensional electron 
layers. 
 
We solve the coupled equations numerically and use the resulting eigenenergies
and eigenfuctions to evaluate the cyclotron effective mass and
transition matrix elements. The equation (\ref{evenodd}) helps to
determine  whether the Landau level belongs to the fan of a ground  
or an excited subband.

The above desribed approach is not appropriate when $B_{\perp}$ is
very small in comparison with $B_{\|}$. In such a case the
quasi-classical description \cite{smrcka,s&t} should be employed.
For small $B_{\perp}$-components of magnetic fields it yields the same
results as the quantum mechanical approach 
but it fails when both $B_{\|}$ and $B_{\perp}$ are strong.
Then the quantum mechanical approach must be used 
and the interpretation of cyclotron resonance experiments becomes 
more complicated as cyclotron effective masses depend on both $B_{\|}$ 
and $B_{\perp}$.


\section{Results and discussion}
\label{results}

\subsection{Two-dimensional electron single-layers}
\label{singlew}

\enlargethispage{0.5cm}
Within the parabolic-confinement approximation, the 2D electron single-layer 
system decouples into two harmonic oscillators with eigenfrequencies
(\ref{omega12}). The confinement frequency $\Omega$ yields the subband
separation in perpendicular magnetic field.

In tilted magnetic fields, if $B_{\perp}$ is small 
($\omega_{\perp}\ll\Omega$), $\omega_{1}$ is determined mainly 
by $\omega_{\perp}$ and the quantum number $n$ plays the role of 
the Landau band index, while $\omega_{2}$ reduces to $\Omega$ and 
the quantum number $m$ represents the subband index. 
In high perpendicular magnetic fields ($\omega_{\perp}\gg\Omega$) 
the roles of both indices are interchanged.

\begin{figure}[t]
\begin{center}
\begin{tabular}{cc}
\hbox{
\includegraphics[scale=0.35]{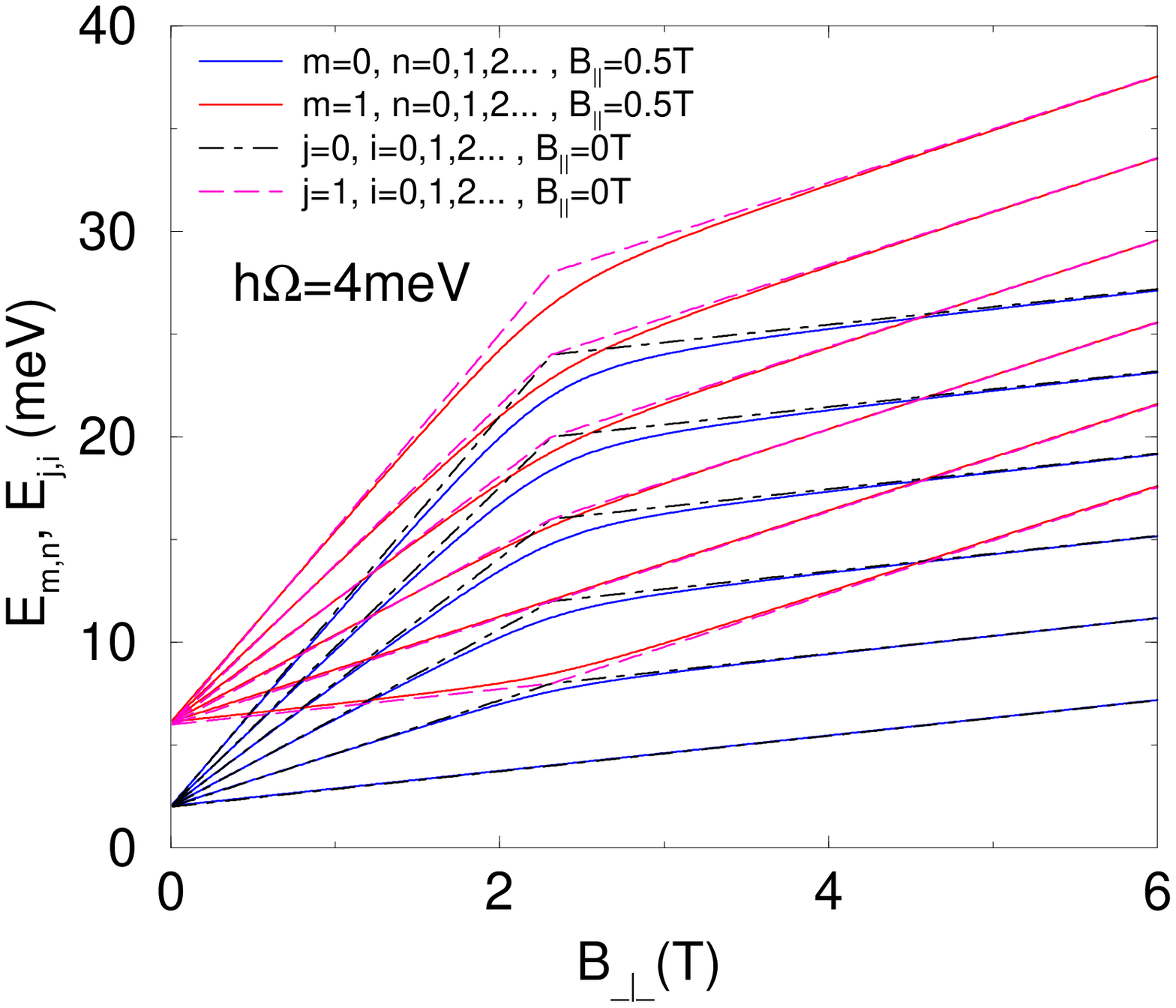}} &
\hbox{
\includegraphics[scale=0.35]{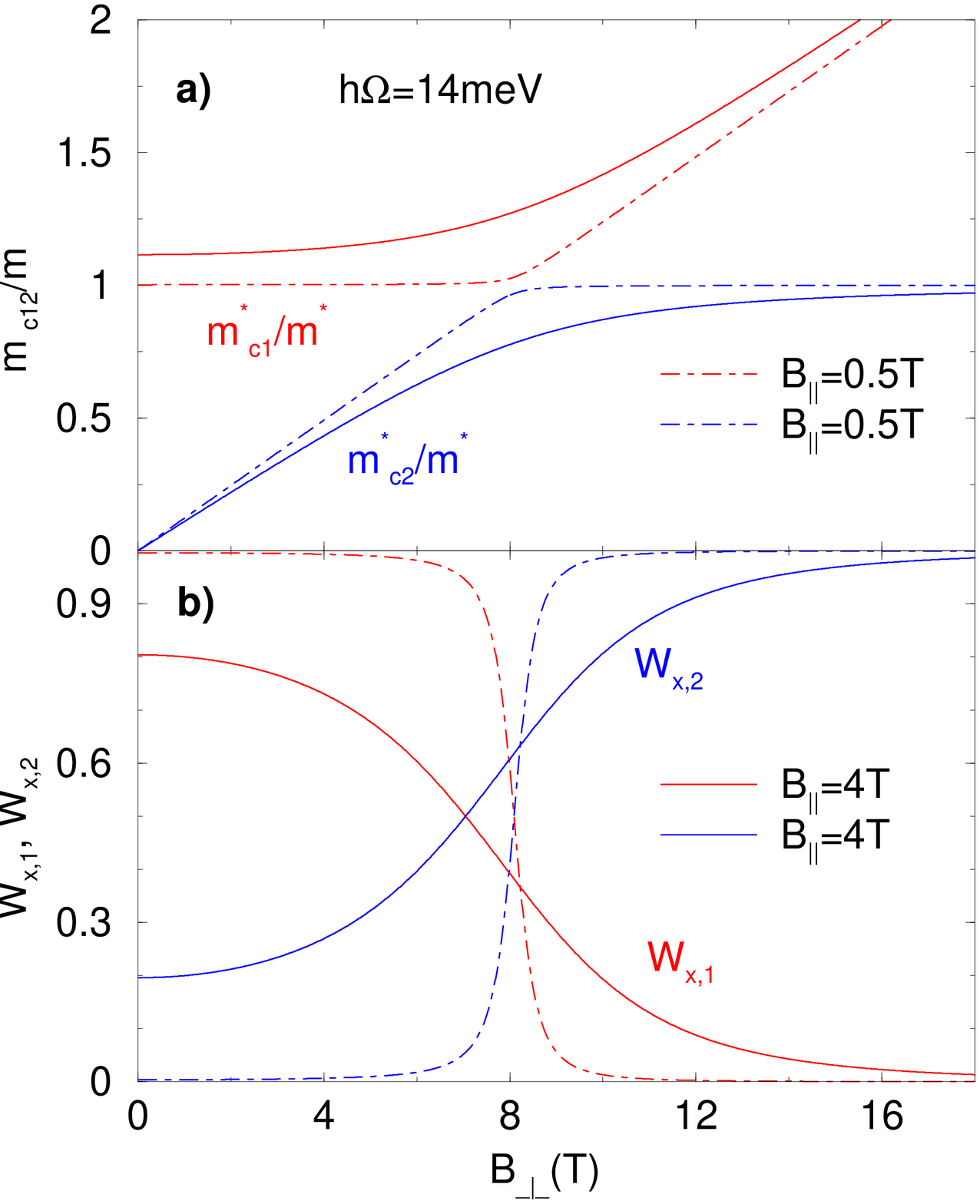}}
\end{tabular}
\caption{\label{fig01}(left) Fan diagram of eigenenergies against the 
$B_{\perp}$\discretionary{-}{-}{-}component of magnetic field 
corresponding to subband separation energy 
$\hbar\Omega=4$~meV for the model of 2D electron single-layer systems. 
Solid curves represent the case of fixed $B_{\|}=0.5$~T. 
Dotted curves correspond to $B_{\|}=0~$T.}
\caption{\label{fig02}(right) The 
$B_{\perp}$\discretionary{-}{-}{-}dependence 
\textbf{\textit{a)}} of the $m^{\ast}_{c1,2}$, and \textbf{\textit{b)}} 
of weights $W_{x,1}$, $W_{x,2}$ of allowed optical transitions between 
Landau levels in 2D electron single-layer systems corresponding 
to $B_{\|}=0.5$~T (dotted curves) and $B_{\|}=4~$T (solid curves). All curves 
are calculated for the narrow quantum well with the confinement 
energy $\hbar\Omega=14$~meV.} 
\end{center}
\end{figure}

The $B_{\perp}$-dependence of eigenenergies $E_{0,n}$ and  
$E_{1,n}$, $n=0,1,2\ldots$ is shown in figure~\ref{fig01}. The in-plane 
component of magnetic field increases the subband separation and changes 
the Landau-level structure. 
In tilted magnetic fields Landau-level mixing arises and not only 
intrasubband transitions between adjacent Landau levels of a single 
subband but also intersubband transitions between Landau levels 
with the same index and from different subbands are allowed.    
Figure~\ref{fig02} presents the field dependence of two different types of effective 
cyclotron masses associated with the eigenfrequencies $\omega_{1}$, 
$\omega_{2}$ and of weights of the intrasubband ($W_{x,1}$) and the 
intersubband ($W_{x,2}$) transitions are governed by the transition 
matrix elements which, similarly as $m^*_c$ itself,
strongly depend on both $B_{\|}$ and $B_{\perp}$.    
%
%

\begin{figure}[t]
\begin{center}
\begin{tabular}{cc}
\hbox{
\includegraphics[scale=0.4]{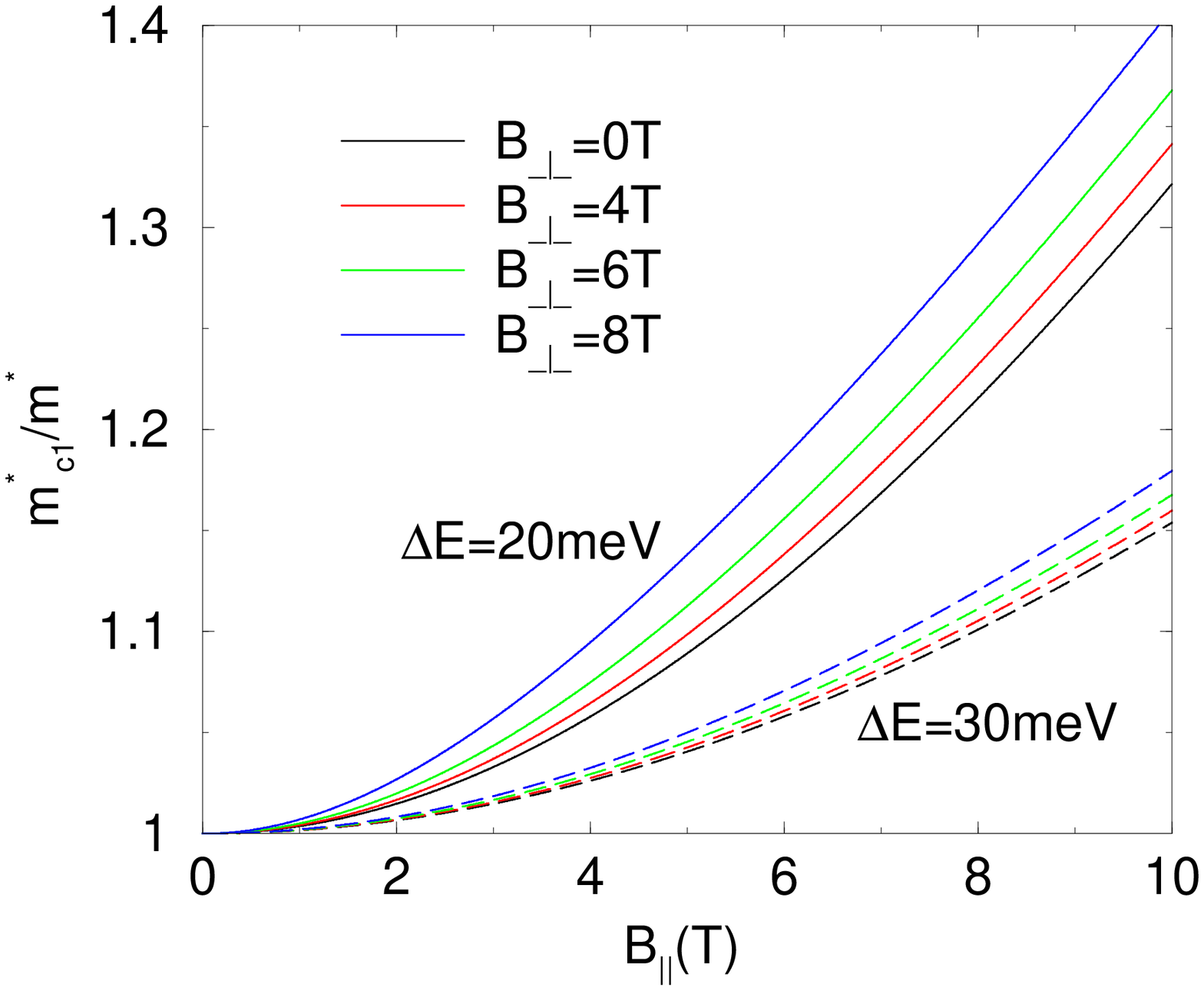}} &
\hbox{
\includegraphics[scale=0.4]{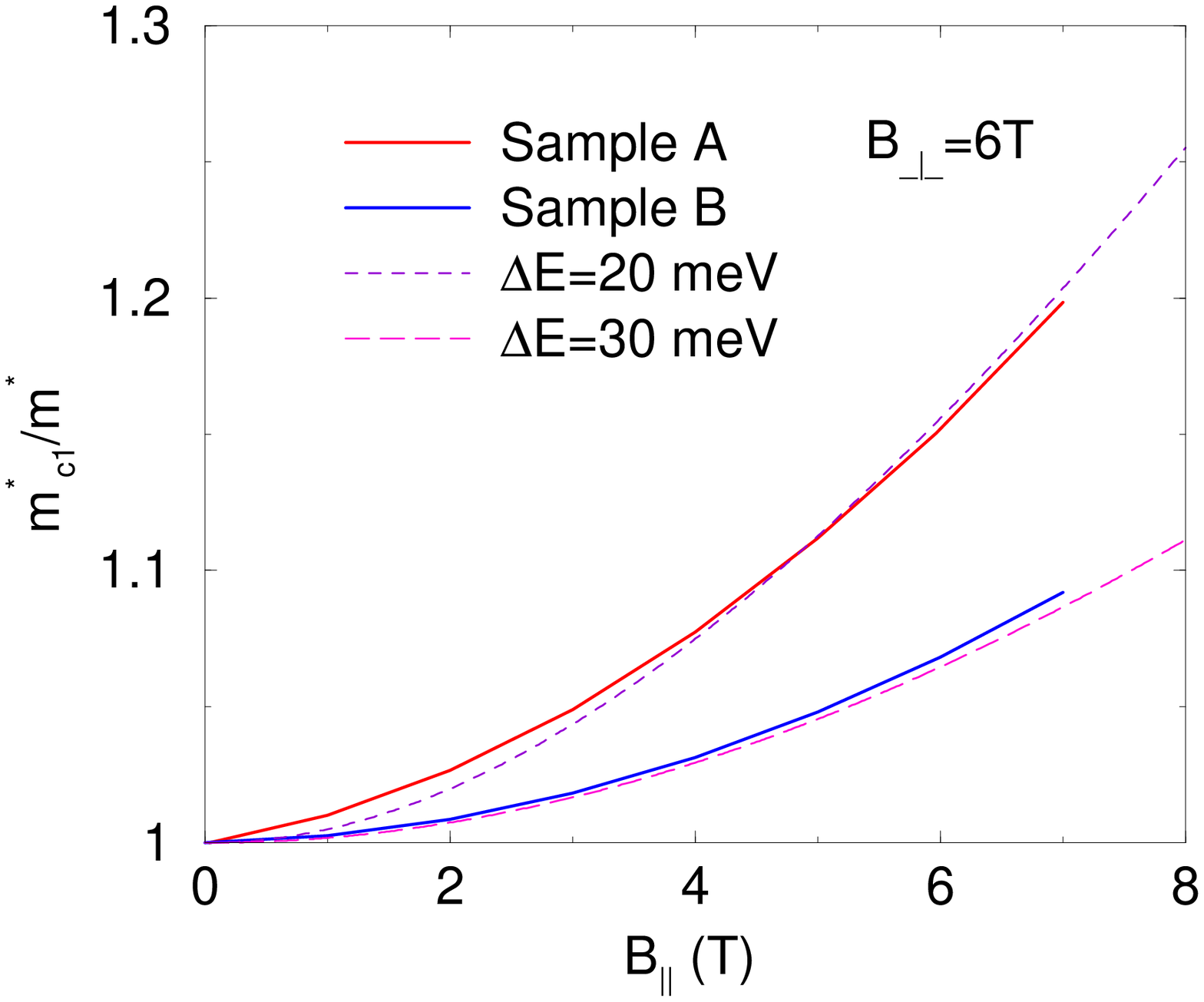}}
\end{tabular}
\caption{\label{fig03}(left) The $B_{\|}$-dependence of $m^{\ast}_{c1}$ 
for different fixed $B_{\perp}$ components of magnetic field and for 
different quantum well widths characterized by the subband 
separation parameter $\Delta E = E_{1,n}-E_{0,n} = \hbar\Omega$.}
\caption{\label{fig04}(right) The $B_{\|}$ dependence of $m^{\ast}_{c1}$ at 
$B_{\perp}=6$~T. Theoretical curves are shown in comparison with 
experimental results of S.~Takaoka~\cite{takaoka} on samples 
with different electron concentrations 
($2.0\times 10^{11}~cm^{-2}$ in sample A and 
$3.0\times 10^{11}~cm^{-2}$ in sample B).}
\end{center}
\end{figure}

If 2D electron systems have only one occupied 
subband, various electron concentrations can be modelled 
by changing $\Omega$, the higher the concentration, the narrower 
the well. We have found that wider parabolic wells 
are more sensitive to the magnitude of both magnetic field components 
than narrower (figure~\ref{fig03}) and have shown (figure~\ref{fig04})
that our model can reasonably fit the $B_{\|}$- and
$B_{\perp}$\discretionary{-}{-}{-}dependence 
of experimental data~\cite{takaoka}.  
%
%

\enlargethispage{0.5cm}
A convenient way to measure the cyclotron resonance in 2D electron systems is 
FIR optical magneto-absorption measured by Fourier transform spectrometer.
The absorption process
is accompanied by electron excitation from an occupied to an
unoccupied level whereby the energy difference is an integer
multiple of $\hbar\omega_{\perp c} = \hbar |e| B_{\perp}/m^*_c$.
At radiation energies equal to these differences the measured 
magnetoconductivity exhibits maxima.

\begin{figure}[t]
\begin{center}
\begin{tabular}{cc}
\hbox{
\includegraphics[scale=0.4]{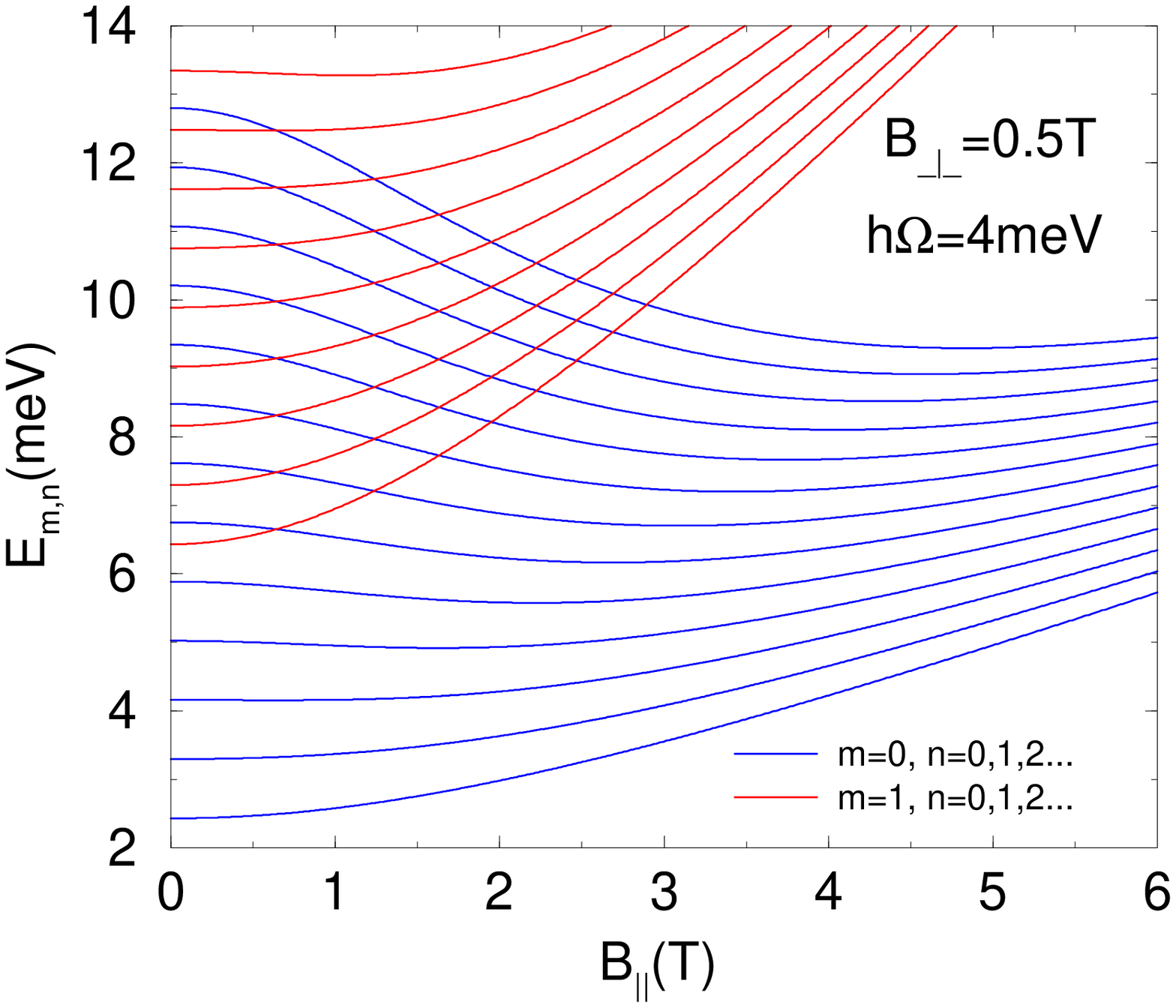}} &
\hbox{
\includegraphics[scale=0.4]{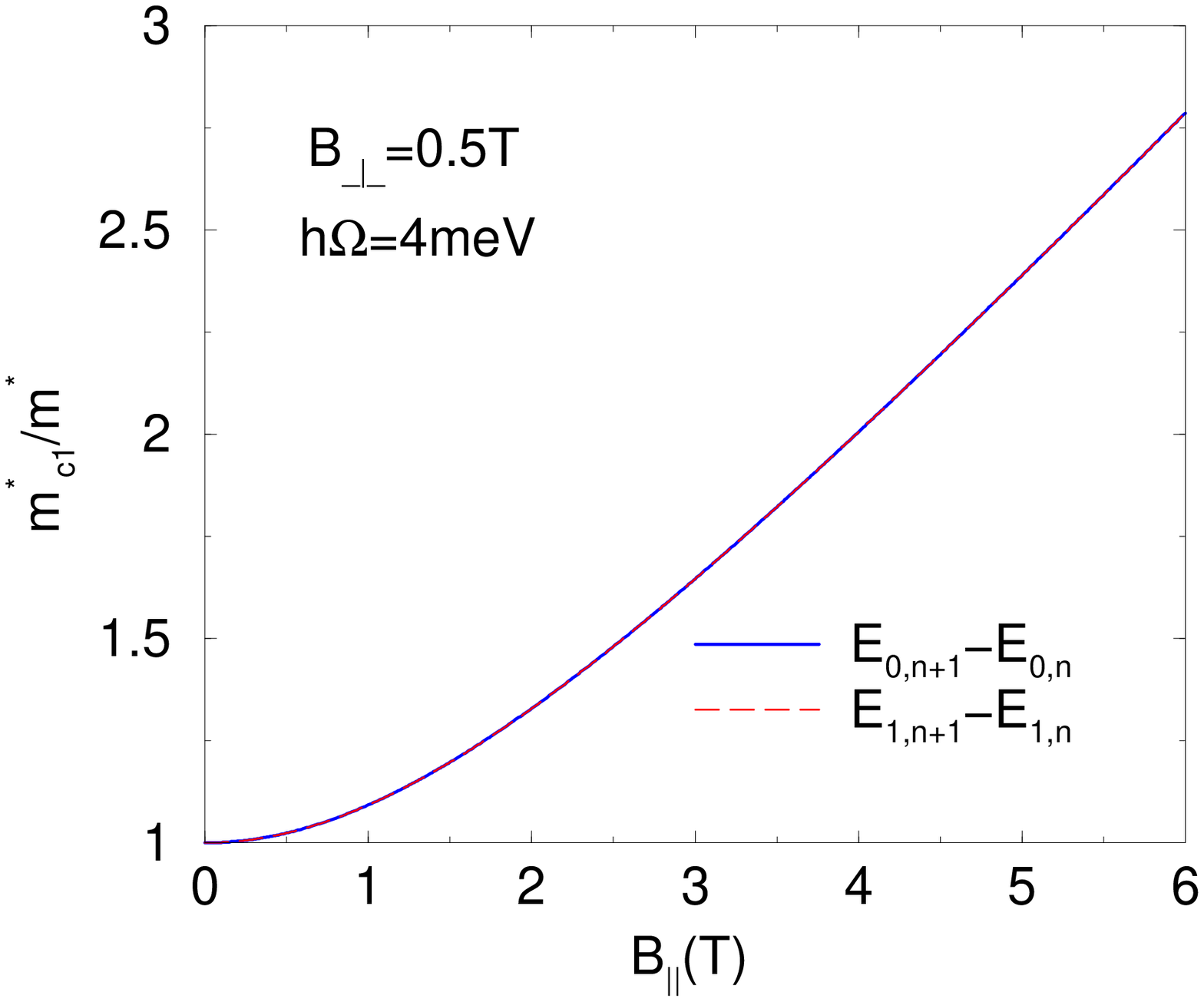}}
\end{tabular}
\caption{\label{fig05}(left) Fan diagram of eigenenergies against the in-plane 
$B_{\|}$ component of magnetic field at fixed $B_{\perp}=0.5T$ corresponding 
to the energy of subband separation $\hbar\Omega=4meV$ for the model of 
2D electron single-layer systems.}
\caption{\label{fig06}(right) Cyclotron effective masses $m^{\ast}_{c1}$ 
calculated from the difference between Landau levels of the eigenenergy 
spectrum (figure~\ref{fig05}) for two types of transitions 
$\delta m=0$ $\delta n=\pm 1$ and 
$\delta m=\pm 1$ $\delta n=0$ as a function of the in-plane magnetic field.}
\end{center}
\end{figure}

The effective cyclotron mass can be calculated from 
the separation energy of Landau levels corresponding to allowed intrasubband 
or intersubband transitions. For fixed 
$B_{\perp}$ and $B_{\|}$ Landau levels  
are equidistant (figure~\ref{fig05}) and cyclotron effective masses connected 
with different transitions are identical (figure~\ref{fig06}).     

%
%

\subsection{Two-dimensional electron double-layers}
\label{doublew}
The coupling between quantum wells due to electron tunneling through 
the barrier removes the degeneracy of the energy 
spectrum and the lowest bound states of individual quantum wells form
symmetric and antisymmetric pairs. In the perpendicular magnetic field 
the separation between symmetric and 
antisymmetric states is characterised by the interwell hopping 
energy $2|t|$. 
Results for 2D electron bilayers were calculated numerically.
As follows from the Kubo formula (\ref{kubo}), the diagonal components
of the magnetoconductivity can be expressed in terms of a sum 
of dimensionless oscillator strengths $f_{\alpha , 1}$, 
$f_{\alpha , 2}$~\cite{davies} where the index $\alpha$ denotes $x$ and $y$ 
as in the case of single-layer, and the indices 1 and 2 indicate 
intrasubband and intersubband transitions, respectively. 
Due to system symmetry, selection rules hold~\cite{davies} 
and only transitions with nonzero oscillatory strength are allowed. 
These include intrasubband transitions 
$\delta m=0$, $\delta n=\pm 1 $ 
and intersubband transitions $\delta m=\pm 1$, $\delta n = 0$.

\begin{figure}[b]
\begin{center}
\begin{tabular}{cc}
\hbox{
\includegraphics[scale=0.4]{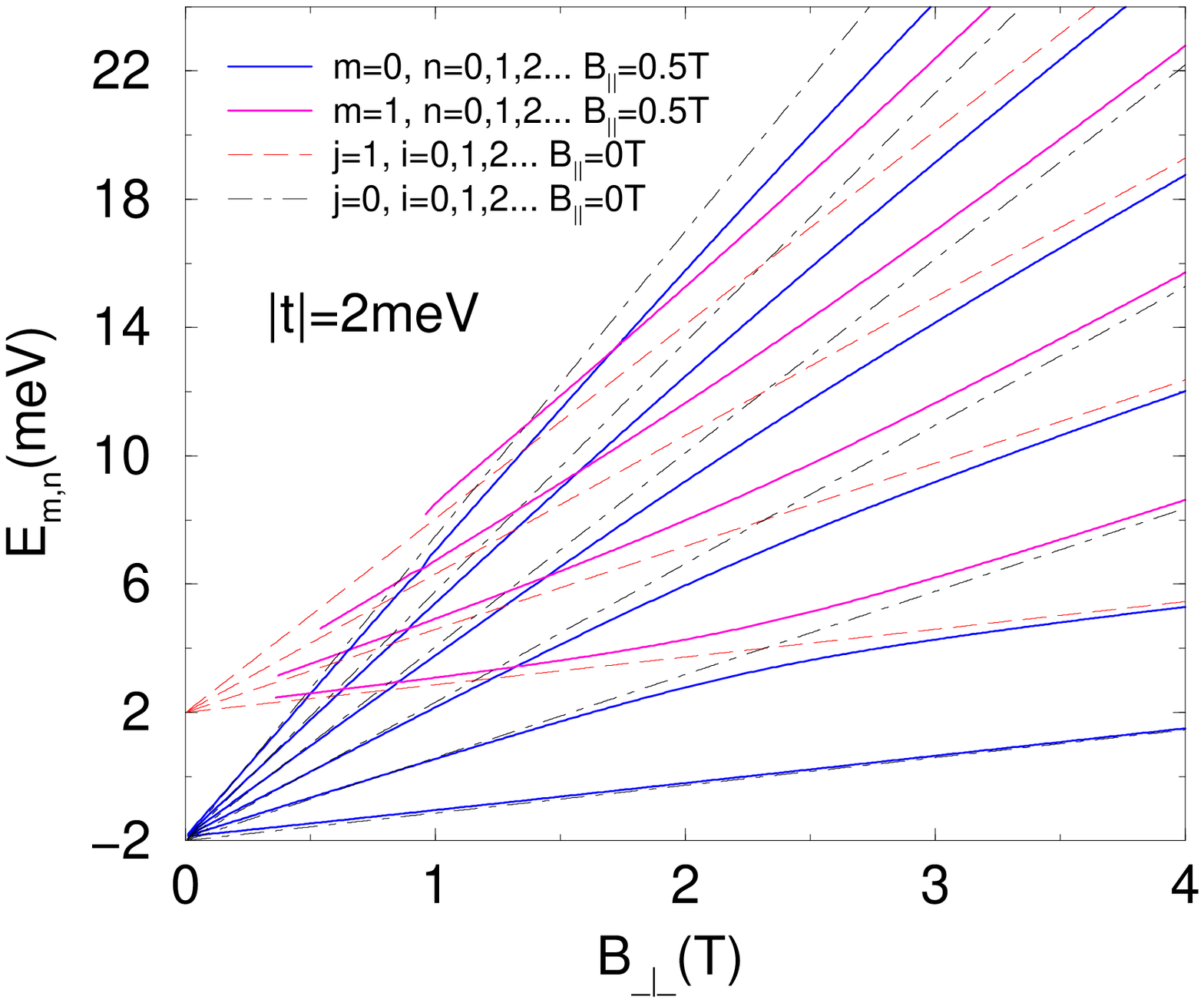}} &
\hbox{
\includegraphics[scale=0.4]{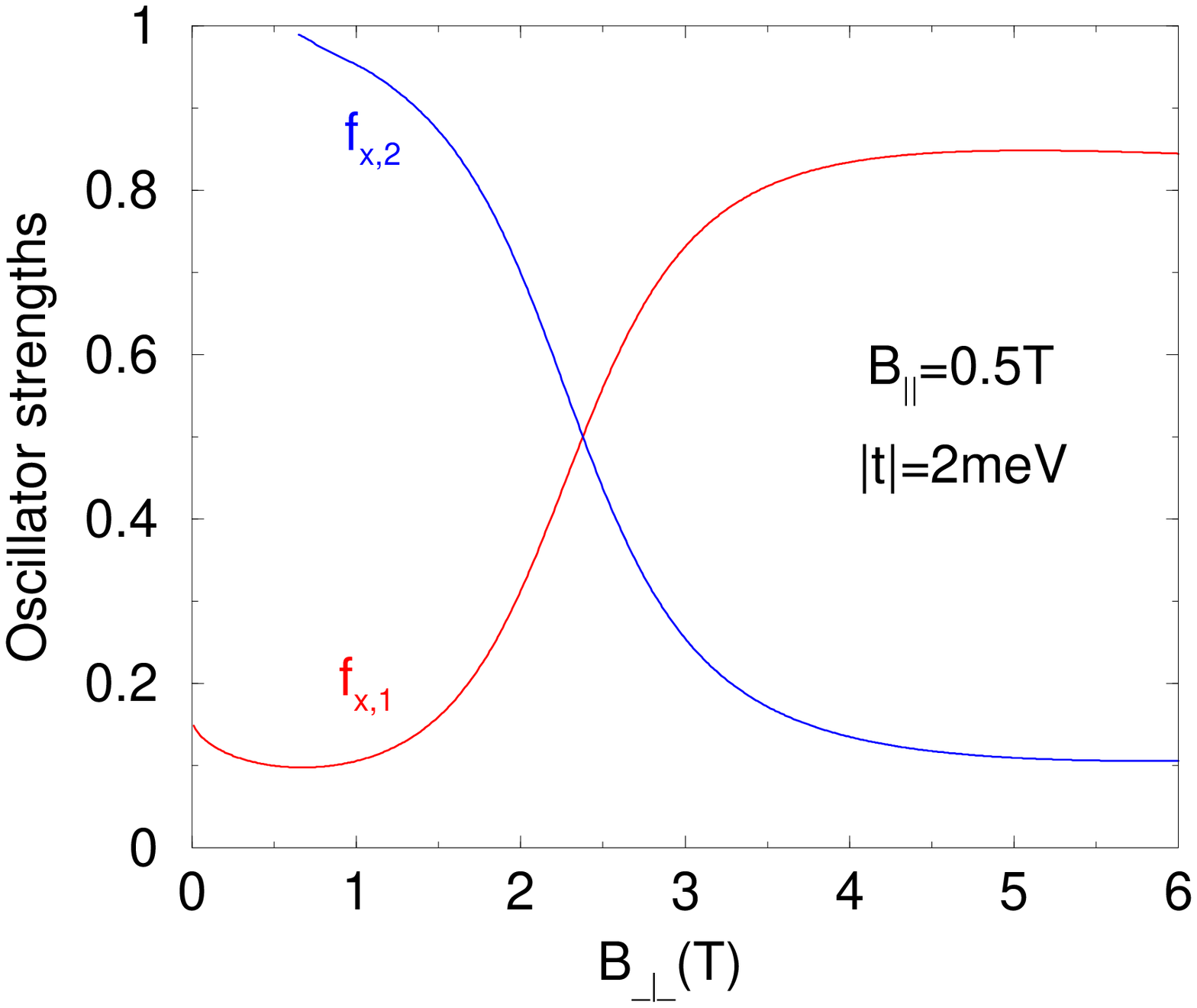}}
\end{tabular}
\caption{\label{fig07}(left) Eigenenergies in the 2D electron bilayer 
systems as a function of $B_{\perp}$.
Solid curves correspond to tilted 
magnetic fields with fixed $B_{\|}=0.5T$ with an anticrossing due 
to subband-Landau-level coupling $|t|=2meV$. 
Blue lines represent the bonding  and red lines the antibonding 
Landau-level subband. Dotted curves represent the case where
$B_{\|}=0T$.}
\caption{\label{fig08}(right) The $B_{\perp}$-dependence  
of oscillator strengths $f_{x,1}$, $f_{x,2}$ of allowed optical 
transitions between Landau levels in 2D electron double-layer 
systems corresponding to fixed $B_{\|}=0.5T$ and $|t|=2meV$. 
Index $1$ determines intrasubband transitions between adjacent 
Landau levels $m=0, n=0$ and $m=0, n=1$ from bonding subband. 
Index $2$ denotes intersubband transitions between Landau levels 
$m=0, n=0$ from bonding subband and $m=1, n=0$ from antibonding subband.}
\end{center}
\end{figure}

In what follows we concentrate on the lowest transitions from $n=0$ which 
are more pronounced than all other channels.
Figures~\ref{fig07} and \ref{fig08} show $B_{\perp}$-dependence of 
eigenenergies calculated for a symmetrical double well 
structure and of oscillator strengths corresponding to the 
strongest permitted transitions in given system.  

\begin{figure}[t]
\begin{center}
\begin{tabular}{cc}
\hbox{
\includegraphics[scale=0.4]{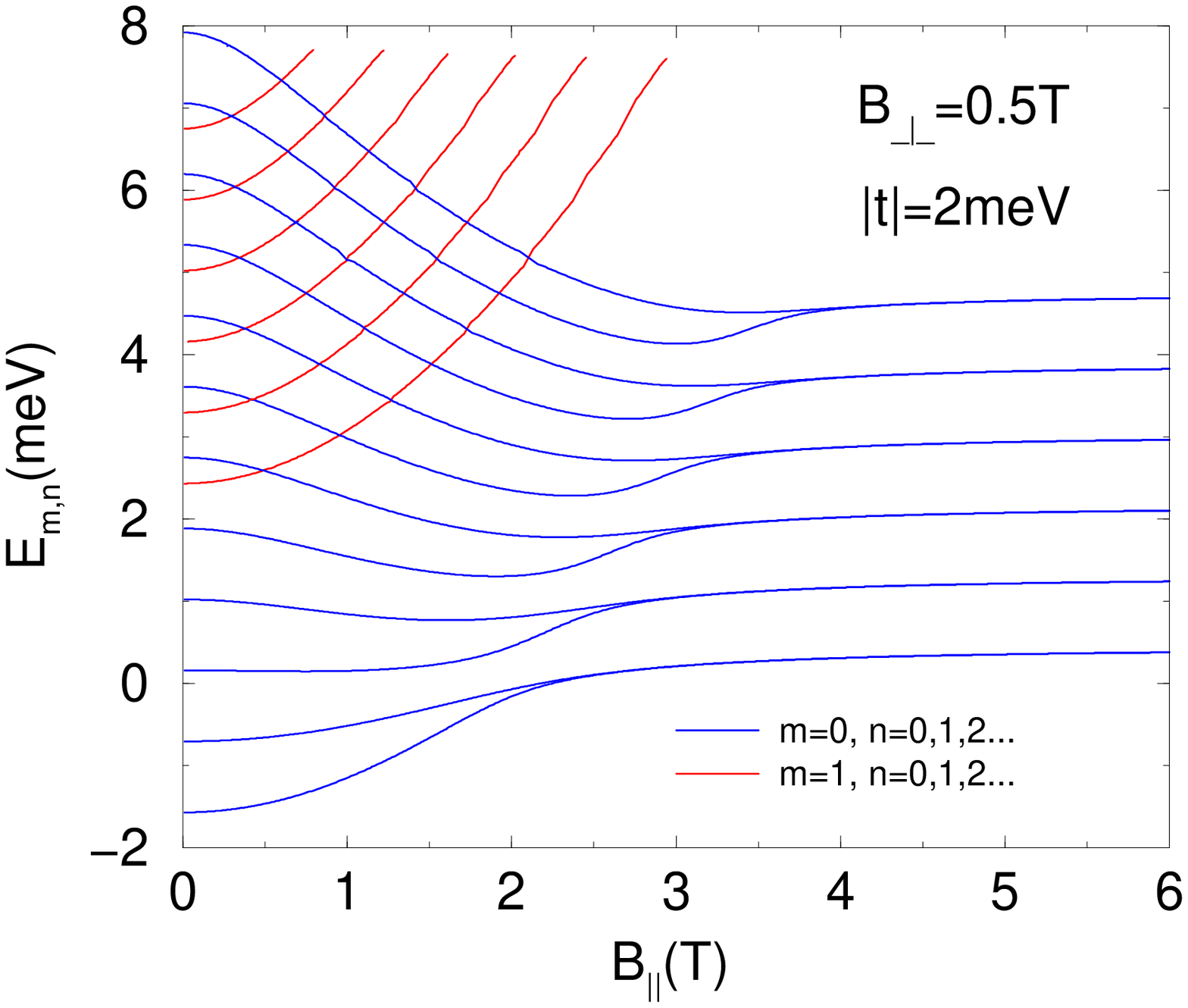}} &
\hbox{
\includegraphics[scale=0.4]{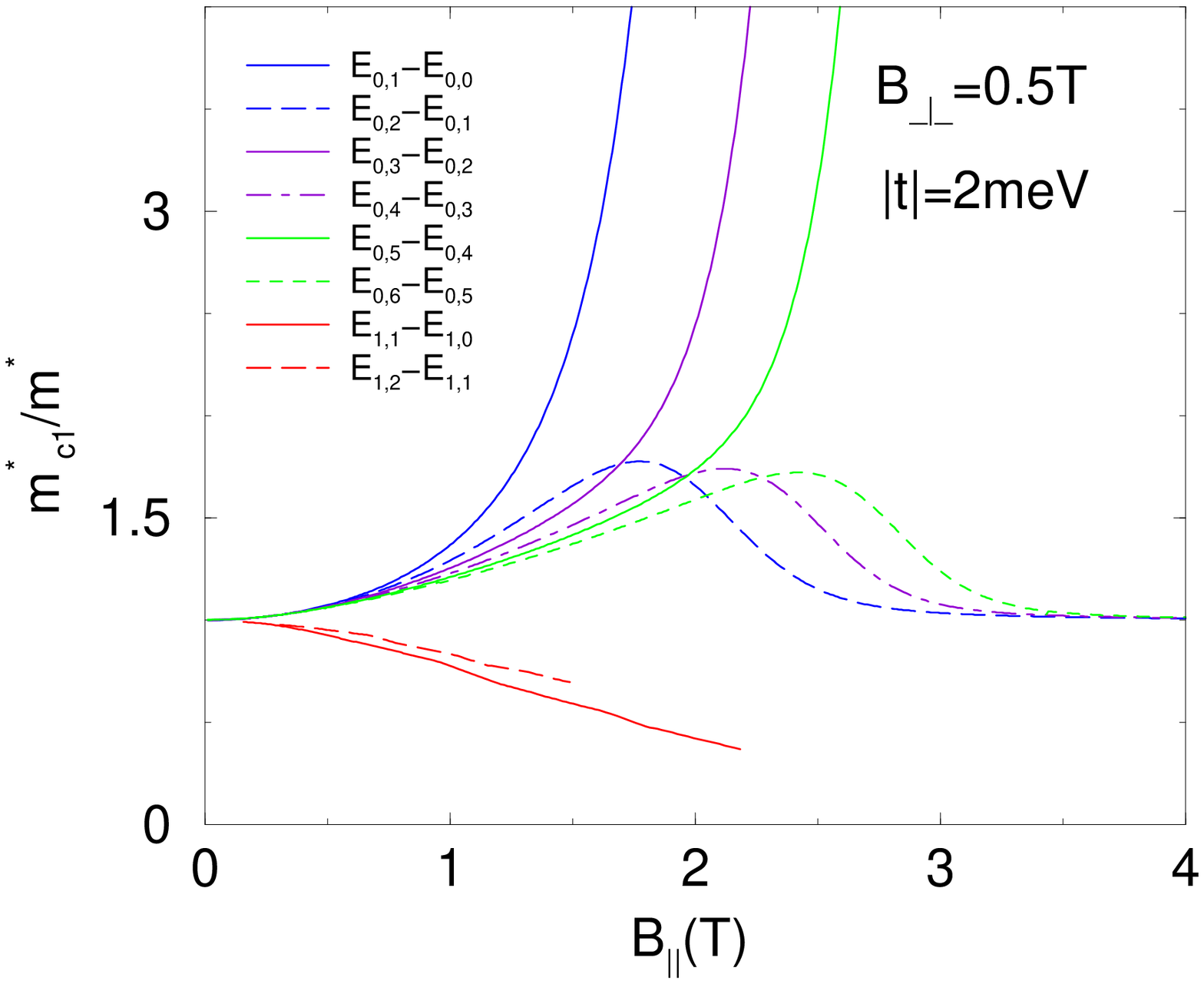}}
\end{tabular}
\caption{\label{fig09}(left) Eigenenergies of the 2D electron bilayer 
systems as a function of $B_{\|}$ at fixed $B_{\perp}=0.5$~T and $|t|=2meV$. 
Blue and red lines denote Landau levels from bonding and antibonding
subband, respectively.}
\caption{\label{fig10}(right) Cyclotron effective masses $m^{\ast}_{c1}$ 
calculated from energy difference between adjacent Landau levels of 
the band structure spectrum (figure~\ref{fig09}) as a function 
of the in-plane magnetic field.}
\end{center}
\end{figure}

The Landau levels with high quantum numbers can be 
described both quasi-classically~\cite{smrcka,s&t} 
and quantum-mechanically. Using our
quantum-mechanical approximation for the case of slightly tilted field 
we have calculated spectrum of eigenenergies. With increasing in-plane 
component of magnetic field due to the process of electron transfer from 
the higher occupied antibonding subband to the lower bonding subband 
the separation energy between bonding and antibonding subbands grows 
until the antibonding subband is completely depopulated. Further growth of 
$B_{\|}$ leads to the degeneracy of Landau levels from the bonding subband
as a result of the process is accompanied by complete suppression of 
interwell tunneling when electrons move in one of individual wells. 
Figure~\ref{fig10} shows the field-dependence of cyclotron effective masses 
calculated from eigenenergy spectrum presented on figure~\ref{fig09}. 
The effective cyclotron masses corresponding to the highest antibonding 
subband decrease with increasing of the in-plane field, while cyclotron 
masses corresponding to the lowest bonding subband increase. 
The positions of singularities of cyclotron masses corresponding to the 
lowest bonding subband are related to the values of $B_{\|}$ 
at which coupled double quantum wells move to two decoupled electron  
layers. As the number of occupied levels depends on the Fermi energy, 
the increase in the electron concentration  
shifts the singularity position to the side of larger in-plane fields.

\begin{figure}[t]
\begin{center}
\includegraphics[angle=-90,scale=0.5]{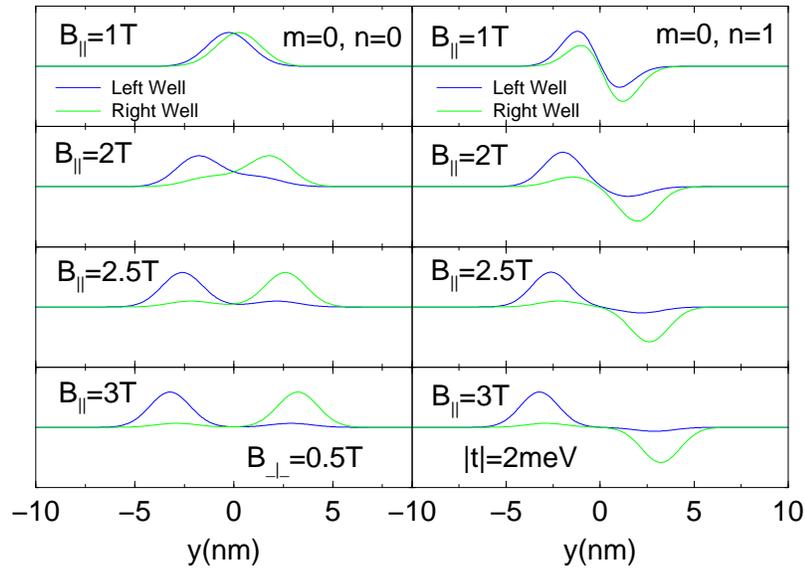}
\caption{\label{fig11}Wave functions of even and odd states of two 
symmetric coupled quantum wells with the coupling parameter $|t|=2meV$ 
for fixed $B_{\perp}=0.5T$ and for a set of $B_{\|}$-components.}
\end{center}
\end{figure}

The field-induced electron-layer decoupling process, shown in
figure~\ref{fig11}, illustrates the aforementioned explanation. 
Wave function components of individual layers $\varphi_{L}(y)$, 
$\varphi_{R}(y)$ are shifted by the distance 
$y_{0} = d B_{\|}/2 B_{\perp}$ where $d$ is 
the distance between the potential minima in the left and the right
well on the z-axis. 
The increase of the in-plane field is followed by a separation growth between 
$\varphi_{L}(y)$ and $\varphi_{R}(y)$. Moreover, with the increase of $B_{\|}$ 
wave function components of higher odd states loose their nodes and 
tend to coincide with the nodeless wave function components 
of lower even states. 

\section{Conclusion}
\label{concl}
To summarize, we have performed 
the theoretical study of the cyclotron resonance in 2D single-layers 
and double-layers subject to magnetic fields of general orientation 
to explain existing experimental results . 
For strong $B_{\|}$ and $B_{\perp}$ the quantum-mechanical approach 
must be involved and $m^*_c$ becomes a function of both components. 
We have calculated the field \discretionary{-}{-}{-} dependence 
of the energy levels and transitions matrix elements between them.
The Kubo formula was employed to 
derive diagonal components of magneto-conductivity tensor for both 
systems. The resulting conductivities are linear combinations of 
two 2D conductivities connected with intrasubband and intersubband 
transitions. 
Conclusions obtained for the model of the parabolic quantum well and 
for the model of coupled double-quantum wells are analogous. 
All results are obtained in a good qualitative agreement with 
experimental data. 

\section{Acknowledgements}
\label{acknowl}
This work has been supported by the Grant Agency of the Czech Republic 
under Grant No $202/01/0754$. The experimental curves in figure 4 are 
presented with kind permission of S.~Takaoka.

\end{document}